\documentclass[10pt,final]{IEEEtran}

\usepackage{cite}
\usepackage{amssymb}
\usepackage{setspace,ulem}

\usepackage{color}   
\usepackage[english]{babel}

\usepackage[pdftex]{graphicx}
\usepackage[cmex10]{amsmath}
\usepackage{algorithmic}

\usepackage{multirow}

\begin{document}

\title{Robust Sparse Blind Source Separation}

\author{C\'{e}cile Chenot, J\'{e}r\^{o}me Bobin and J\'{e}remy Rapin
\thanks{C.C, J. B and J.R with CEA, IRFU, Service d'Astrophysique - SEDI, 91191 Gif-sur-Yvette Cedex, France.}
\thanks{This work was (partially) funded by the PHySIS project, contract no. 640174, within the H2020 Framework Program of the European Commission.}}
\maketitle

\begin{abstract}

Blind Source Separation is a widely used technique to analyze multichannel data. In many real-world applications, its results can be significantly hampered by the presence of unknown outliers. In this paper, a novel algorithm coined rGMCA (robust Generalized Morphological Component Analysis) is introduced to retrieve sparse sources in the presence of outliers. 
It explicitly estimates the sources, the mixing matrix, and the outliers. It also takes advantage of the estimation of the outliers to further implement a weighting scheme, which provides a highly robust separation procedure. Numerical experiments demonstrate the efficiency of rGMCA to estimate the mixing matrix in comparison with standard BSS techniques.
 
\end{abstract}


\IEEEpeerreviewmaketitle

\section{Introduction}

\newcommand{\Sources}{\mathbf{S}}
\newcommand{\matM}{\mathbf{M}}
\newcommand{\matZ}{\mathbf{Z}}
\newcommand{\Weight}{\mathbf{W}}
\newcommand{\Mesures}{\mathbf{X}}
\newcommand{\Gaussien}{\mathbf{N}}
\newcommand{\Amel}{\mathbf{A}}
\newcommand{\AmelEst}{\tilde{\Amel}}
\newcommand{\Outliers}{\mathbf{O}}
\newcommand{\OutliersEst}{\tilde{\mathbf{O}}}
\newcommand{\SourcesEst}{\tilde{\mathbf{S}}}
\newcommand{\WeightEst}{\tilde{\mathbf{W}}}
\newcommand{\lambdaEst}{\tilde{\lambda}}
\newcommand{\alphaEst}{\tilde{\alpha}}
\newcommand{\nbreMesures}{m}
\newcommand{\nbreSources}{n}
\newcommand{\nbreSamples}{t}
\newcommand{\norm}[2]{\left\Vert  {#1} \right\Vert _{#2}}
\newcommand{\vecBeta}{\mathbf{\beta}}
\newcommand{\vecLambda}{\mathbf{\lambda}}
\newcommand{\softT}[2]{\mathcal{S}_{#1}\hspace{-1.0mm}\left(\hspace{-0.5mm} #2 \hspace{-0.5mm}\right)}
\newcommand{\argmin}{\operatorname*{arg min}}
\newcommand{\Divergence}[2]{\mathcal{D}\left( #1 \Arrowvert #2 \right)}
\newcommand{\penFunc}[1]{\mathcal{J}\left( #1 \right)}

\IEEEPARstart{T}{he} advances in multichannel technologies are exploited in various fields such as astrophysics \cite{PR1_LGMCA} or hyperspectral remote-sensing \cite{Ma2014SigProcPerspHypUnm}. This has generated a considerable interest for methods able to extract the relevant information from these specific data. Blind Source Separation (BSS) is one of them. In this context, the $\nbreMesures$ noisy observations $\left\lbrace \Mesures_i\right\rbrace _{i=1..m}$ are assumed to be the linear mixture of $\nbreSources\leq \nbreMesures$ sources $\left\lbrace \Sources_j\right\rbrace _{j=1..n}$ with $\nbreSamples>\nbreMesures$ samples. The presence of instrumental noise and model imperfections is usually admitted and represented by a Gaussian noise $\Gaussien$. The BSS techniques aim to estimate $\Amel$ and $\Sources$ from $\Mesures$. This problem is well-known to be ill-posed as the number of solutions is infinite. This requires imposing further prior information to recover the original sources such as the statistical independence of the sources (ICA methods \cite{comon2010handbook} and references therein) or the sparsity of $\Sources$ \cite{zibulevsky2001Blind,bobin2007sparsity,ica:cichocki}.\\

Most of the techniques in BSS are highly sensitive to the presence of spurious outliers while these ones are frequent in real-world data \cite{gadhok2005rotation}. Indeed in many applications, the data are additionally corrupted by a few, and therefore sparse, large errors that cannot be modeled by standard additive Gaussian noise: this includes stripping noise \cite{torres2001wavelet}, impulsive noise \cite{chen2013impulsive}, glitches \cite{ordenovic2008detection} to name only a few. 
In the next, we will consider that the observations can be expressed as:
$$\Mesures=\Amel\Sources+\Outliers+\Gaussien,$$ where $\Mesures\in \mathbf{R}^{\nbreMesures\times\nbreSamples}
$ stands for the observations, $\Amel\in \mathbf{R}^{\nbreMesures\times\nbreSources}$ the mixing matrix, $\Sources\in \mathbf{R}^{\nbreSources\times\nbreSamples}$ the sources, $\Outliers\in \mathbf{R}^{\nbreMesures\times\nbreSamples}$ the outliers, and $\Gaussien\in \mathbf{R}^{\nbreMesures\times\nbreSources}$ the Gaussian noise. \\

To be best of our knowledge, only few blind source separation methods have been proposed for coping with outliers. So far, the state-of-the-art techniques can be divided into three groups. The first approach consists in replacing the data fidelity term in the standard methods by some divergence that provides more robustness to the presence of outliers, without estimating them explicitly \cite{hamza2006reconstruction}, \cite{kong2011robust}. In particular, the use of the $\beta$-divergence, which is a generalization of the Kullback-Leibler divergence, has received a growing attention. In \cite{mihoko2002robust}, a general BSS method using statistical independence and minimizing the $\beta$-divergence has been proposed. The numerical experiments in \cite{mihoko2002robust} show that by finding an appropriate value of $\beta$, the estimation of $\Amel$ becomes quite robust to the outliers. However, the selection of $\beta$ is a challenging task in practice and this method only estimates the mixing matrix: the sources and the outliers remain unmixed.\\
The second scheme proceeds in a two-steps strategy that consists in: i) pre-processing the data to discard the outliers, and ii) performing source separation on the pre-processed estimate of the data $\Amel\Sources$. Several outlier removal methods that exploit the low-rank property ($\nbreMesures\gg\nbreSources$) of $\Amel\Sources$ have been recently proposed, especially in hyperspectral imaging \cite{li2014denoising,zhang2015}. Indeed, it has been shown that it is possible to separate $\Amel\Sources$ and $\Outliers$ with a high accuracy if $\Amel\Sources$ has low-rank and if the outliers are sparsely distributed and additionally assumed to be in general position (they do not cluster in specific directions) \cite[PCP]{candes2011robust}. However, these assumptions are generally not met in practice, in particular when the number of observations is close to the number of sources. The major drawback of this approach is that an inaccurate estimation of the mixture $\Amel\Sources$ will very likely hamper the separation process. Moreover, the parameters of these methods need to be tuned to return a sufficiently accurate estimation of $\Amel\Sources$ prior to performing the separation.\\
Last, the third strategy consists in estimating jointly the sources, the mixing matrix and the outliers \cite{fevotte2014nonlinear,shen2012robust}. This leads to a flexible framework in which the priors on the components can be individually taken into account. In \cite{fevotte2014nonlinear} for example, the authors use the $\beta$-divergence for the data fidelity term $\Mesures-\Amel\Sources-\Outliers$, the $\ell_{2,1}$ norm for $\Outliers$, which corrupts entirely some columns of $\Mesures$, to enforce its sparsity and the positivity prior for $\Amel$ and $\Sources$. This third category has only been developed in the scope of NMF, where further enforcing the positivity of $\Amel$ and $\Sources$ is known to greatly enhance the separation process. Since non-negativity is not always a valid assumption in physical applications ({e.g.} astrophysics \cite{PR1_LGMCA}), neither the mixing matrix nor the sources will be assumed to be non-negative in the next.\\

\underline{\it Contribution:} a novel robust algorithm for sparse BSS is proposed. To the best of our knowledge, no method using sparsity for tackling BSS problems in the presence of outliers has been studied so far. Building upon the sparse BSS algorithm GMCA, our algorithm coined rGMCA (robust GMCA) estimates jointly $\Amel$, $\Sources$ and $\Outliers$ based on the sparsity of the outliers and the sources. Besides, and apart from presuming that the outliers are sparse and in general position, the algorithm makes no assumption about the sparsity pattern of the outliers. This makes the proposed algorithm very general and suitable for various scenarios.\\
The paper is organized as follows: Section II presents the basics of GMCA, Section III introduces rGMCA, and last, numerical experiments are performed to compare rGMCA with standard BSS methods in Section IV.

\section{Sparse BSS}
The sparsity prior has been shown to be an effective separation criterion for BSS \cite{zibulevsky2001Blind,bobin2007sparsity}. Essentially, it assumes that the sources are sparse, which means that they can be described by few coefficients in a given dictionary $\bf \Phi$. In the present paper, we will assume that the sources are sparse in the sample domain: ${\bf \Phi} = {\bf I}$. The morphological diversity property (MDP - see \cite{bobin2007sparsity}) further assumes that the sparse sources do not share their largest entries. This allows to differentiate the sources from their most significant entries in $\bf \Phi$. The algorithm GMCA \cite{bobin2007sparsity} takes advantage of the MDP to seek the sources $\Sources$ and the mixing matrix $\Amel$ from the observations $\Mesures=\Amel\Sources +\Gaussien$ by solving:
 \begin{equation*}\label{mod:GMCA}
 \argmin_{\AmelEst, \SourcesEst}\frac{1}{2}\norm{\Mesures- \AmelEst\SourcesEst }{2}^2 + \sum_{j=1}^{\nbreSources}\lambda_j \norm{\SourcesEst_j}{p},
 \end{equation*}
where the quadratic term is the data fidelity term, which is well suited to deal with Gaussian noise, and the $p$-norm with $p \leq 1$ enforces the sparsity of the recovered $\Sources$. In practice, we customarily choose the convex $\ell_1$ norm. Whereas the above problem is not convex, it can be tackled efficiently with Block Coordinate Relaxation (BCR \cite{tseng2001convergence}) and alternative projected least squares as follows:
\begin{itemize}
	\item Update of $\AmelEst$ for fixed $\SourcesEst$: $$\AmelEst=\argmin_{\AmelEst}\frac{1}{2}\norm{\Mesures- \AmelEst\SourcesEst }{2}^2 ,$$
	obtained with $\AmelEst=\Mesures\SourcesEst^{\dagger}$, where the symbol $.^{\dagger}$ denotes the Moore-Penrose pseudo inverse.
	\item Update of $\SourcesEst$ for fixed $\AmelEst$:$$\argmin_{\SourcesEst}\frac{1}{2}\norm{\Mesures- \AmelEst\SourcesEst }{2}^2 + \sum_{j=1}^{\nbreSources}\lambda_j \norm{\SourcesEst_j}{1},$$
	approximated with the soft-thresholding $\SourcesEst=\softT{\lambda}{\AmelEst^{\dagger}\Mesures}$, where  $\left[\hspace{-0.8mm}\softT{\lambda}{\AmelEst^{\dagger}\Mesures}\hspace{-1mm} \right]_{i,j}\hspace{-5.0mm}=\hspace{-1.4mm} \text{sign}(\AmelEst^{\dagger}\Mesures)_{i,j}\hspace{-1.2mm} \; \odot \; \hspace{-1.2mm}\max (0,\hspace{-0.5mm} \lvert(\AmelEst^{\dagger}\Mesures)_{i,j}\rvert\hspace{-1.2mm} -\hspace{-1.2mm} \lambda_i )$, where the operator $\odot$ stands for the entrywise Hadamard product.
\end{itemize}

In practice, the estimation of the mixing matrix is enhanced by using a decreasing threshold strategy. By starting with a large value of $\lambda$, we only select the largest entries of the sources. 
These large coefficients are weakly influenced by the Gaussian noise and above all, are very likely to belong to only one source (MDP): they are the most discriminant samples for the source separation. The sources are then refined by decreasing the value of $\lambda$ towards $3\sigma$, where $\sigma$ denotes the standard deviation of $\Gaussien$. This final threshold ensures with a high probability that no Gaussian noise contaminates the sources. The noise standard deviation can be estimated using robust empirical estimators such as the median absolute deviation (MAD). Besides, it has been emphasized that the decreasing threshold strategy can prevent GMCA to be trapped into local minima \cite{bobin2007sparsity}. 

\section{Robust GMCA}
The rationale of the proposed separation procedure relies on the difference of structure between the outliers and the term $\Amel\Sources$. Indeed, the outliers are assumed to be in general position while the source contribution $\Amel\Sources$ tends to naturally cluster along the directions described by the columns of the mixing matrix $\Amel$. Following \cite{Rapin_13_SparseandNon}, minimizing the $\ell_1$ norm of the sources tends to favor solutions $\AmelEst$ so that the corresponding $\SourcesEst$ are clustered along such axes. This motivates the use of sparsity to provide an efficient separation scheme. In the following, we therefore introduce a new sparsity-enforcing BSS method based on the GMCA framework.

\subsection{A naive extension}
A straightforward strategy to account for the presence of outliers in the framework of GMCA is done by including an extra sparse term $\Outliers$ enforcing the sparsity of the outliers. This approach, which we coin Naive robust GMCA (NrGMCA), can be formulated as:
\begin{equation*}
\argmin_{\OutliersEst,\AmelEst,\SourcesEst} \frac{1}{2}\norm{\Mesures-\AmelEst\SourcesEst-\OutliersEst}{2}^2  + \sum_{j=1}^{\nbreSources}\lambda_j\norm{\SourcesEst_j}{1} + \alpha\norm{\OutliersEst}{1},
\end{equation*}

where the first term is the data-fitting term, well suited to deal with the Gaussian noise $\Gaussien$, and the two others terms enforce the sparsity of the sources and the outliers. \\
However, this first naive approach cannot handle large outliers, especially if $\Outliers$ have several active entries per row or column. Indeed, in cases where the outliers are the dominant contribution to the data, the MDP does not obviously hold since the largest entries of the data are related to the outliers and not to individual sources. The outliers are then likely to be estimated as sources, misleading the estimation of $\Amel$. 

\subsection{The rGMCA algorithm}

In the presence of large outliers, as the MDP does not hold, discriminating between the $\Outliers$ and $\Amel\Sources$ becomes more challenging and requires at least improving the robustness of the estimation of the mixing matrix against the influence of the outliers. For this purpose, we propose to extend NrGMCA building upon the AMCA algorithm \cite{bobin2014sparsity}. \\

In a different context, the AMCA algorithm extends the GMCA in the special case of sparse and partially correlated sources, where the MDP does not hold either \cite{bobin2014sparsity}. In brief, this method relies on an iterative weighting scheme that penalizes non-discriminant entries of the sources. Inspired by this approach, we propose to implement a similar weighting scheme to penalize samples that are likely to be contaminated with large outliers. In the spirit of AMCA, the influence of the corrupted samples are weakened by using the following weighting scheme in the mixing matrix update stage:

\begin{equation*}
\argmin_{\OutliersEst,\AmelEst,\SourcesEst} \frac{1}{2}\norm{(\Mesures-\AmelEst\SourcesEst-\OutliersEst)\Weight}{2}^2  + \sum_{j=1}^{\nbreSources}\lambda_j\norm{\SourcesEst_j}{1} + \alpha\norm{\OutliersEst}{1},
\end{equation*}
where $\Weight$ is the penalizing diagonal matrix of size $\nbreSamples\times\nbreSamples$. The role of the weighting procedure is to penalize the samples of the sources that are likely to be corrupted with outliers. It is therefore natural to define the weights $\Weight$ based on the current estimate of the outlier matrix $\OutliersEst$. In the spirit of \cite{bobin2014sparsity}, an efficient weighting procedure consists in defining the weights based on the sparsity level of the columns of the outlier matrix as follows: $\Weight_{i,i}=\frac{1}{\varepsilon+\norm{\OutliersEst^i}{1}}$, where $\varepsilon$ stands for $\frac{\text{median}_{\arrowvert S_{(i,i)} \arrowvert>0}\arrowvert S_{(i,i)} \arrowvert}{10}$. Subsequently, the penalization of a given data sample will increase with the amplitude of the outliers as well as the number of outliers per data sample.\\
Following the structure of the GMCA algorithm, this problem is solved by using BCR and alternative projected least squares. The structure of the algorithm is presented in Algorithm 1. Instead of estimating the variables one by one, we found that applying GMCA to the current estimate of $\Mesures-\OutliersEst$ and then to estimate the outliers from $\Mesures-\AmelEst\SourcesEst$ provides the most effective estimation procedure. Indeed, jointly re-estimating  $\AmelEst$ and $\SourcesEst$ from a new estimate of the outlier-cleaned data is more likely to limit the impact of remaining outliers on the source estimation.\\

\subsection{Choice of parameters and initialization}
The large outliers are the most damaging as they can severely mislead the estimation of $\AmelEst$ if they are not estimated as outliers from the start. That is why, the algorithms NrGMCA and rGMCA start by estimating the largest values of $\Mesures$ as outliers. On the other hand, the orientation of the mixing matrix $\AmelEst$, which is initialized as a random matrix whose columns are normalized and entries are Gaussian, is deduced from the remaining large outliers cleaned data $\Mesures-\OutliersEst$: our first estimation of $\OutliersEst$ should not be too conservative to keep the clustering aspect of $\Mesures-\OutliersEst$. For this purpose, we propose to estimate $\OutliersEst$ with a soft-thresholding at the value $\alphaEst^0=\text{median}_{\lvert \Mesures \rvert>3\sigma}\lvert\Mesures\rvert$. Then, similarly to GMCA, the sources are estimated as being the largest components in $\AmelEst^{\dagger}(\Mesures-\OutliersEst)$, determining the corresponding $\lambdaEst^0$. Furthermore, this initial value $\lambdaEst^{0}$ is fixed at the beginning of the algorithm. By keeping a large value of $\lambdaEst^{0}$, we minimize the risk to propagate the errors since only few coefficients from the last estimate of $\SourcesEst$ are kept.\\
Then the decreasing threshold strategy used for $\lambda$ in GMCA is similarly kept. Likewise, the parameter $\alpha$ is also decreasing towards $3\sigma$ to refine $\OutliersEst$ without incorporating too many terms of $\Amel\Sources$ or Gaussian noise. 

\begin{figure}[h]
	\begin{tabular}{c}
		\hline
		Algorithm 1: rGMCA\\
		\hline
		\begin{minipage}{\dimexpr\columnwidth-8\fboxsep-8\fboxrule}
			\vspace{+1mm}
			\begin{algorithmic}
		
				\STATE Initialize $\OutliersEst^{0}$, $\SourcesEst^{0}$, $\AmelEst^{0}$, $\WeightEst^{0}$, $\alphaEst^{0}$, and $\lambdaEst^{0}$.
				\WHILE {$k<K$ } 
				{		
					
					\WHILE {$j<J$  	}
					
					{\STATE $ \SourcesEst^{k,j}=\softT{\lambdaEst^{j}}{\left( \AmelEst^{k,j-1}\right)^{\dagger}(\Mesures-\OutliersEst^{k-1})}$
						\STATE 	
						$\AmelEst^{k,j}=(\Mesures-\OutliersEst^{k-1})\WeightEst^{k}(\SourcesEst^{k,j}\WeightEst^{k})^{\dagger}$
						\STATE Decrease $\lambdaEst^{j}$
					}
					\ENDWHILE

						\STATE $\OutliersEst^{k}=\softT{\alphaEst^{k}}{\Mesures-\AmelEst^{k,J-1}\SourcesEst^{k,J-1}}$	
						
						\STATE Update $\WeightEst^{k}$
						
						\STATE Decrease $\alphaEst^{k}$
					}
						\ENDWHILE
				{\RETURN $\OutliersEst^{K-1}$, $\SourcesEst^{K-1,J-1}$, $\AmelEst^{K-1, J-1}$. }
		
			\end{algorithmic}
		\end{minipage}
	\end{tabular}
\end{figure}

\section{Numerical experiments}
In this section, we compare the performances of rGMCA with standard BSS methods: GMCA \cite{bobin2007sparsity}, PCP+GMCA (the outliers are first estimated with PCP and then discarded from $\Mesures$ \cite{candes2011robust}) , the minimization of the $\beta$-divergence with statistical independence prior (implementation from \cite{gadhok2006implementation}) and NrGMCA. The values of the free parameters for PCP ($\lambda$ in \cite{candes2011robust}) and the $\beta$-divergence minimization algorithm ($\beta$) are tuned to return the best results based on several trials. \\
The minimization of the $\beta$-divergence only returns an estimate for $\Amel$, and thus we propose to compare the methods by using a criterion depending only on $\Amel$ \cite{bobin2007sparsity}: $\Delta_A=\frac{\norm{\AmelEst^{\dagger}\Amel-\mathbf{I}}{1}}{\nbreSources^2}$. The median of $\Delta_A$ obtained from 80 Monte-Carlo  simulations and the percentage of these 80 simulations with an error smaller than $5\times 10^{-3}$ (normalized to represent a probability of success) are represented in the following figures. The probability of success is an interesting indicator of the robustness of the algorithms, showing whether they reliably perform well.\\
Last, the outliers are composed of both broadly distributed errors and entirely corrupted columns (anomalies appearing at a same position in each observation) with random amplitudes, in order to cover a large variety of noise patterns.

\subsection*{Influence of the amplitude of the outliers}
\begin{figure}[h]
	\centering
	\includegraphics[width=0.40\textwidth]{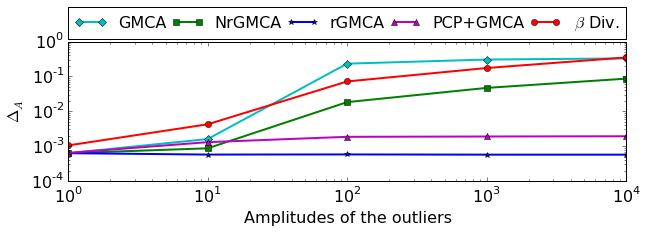}
	\caption{$\Delta_A$ versus the amplitudes of the outliers.}
	\label{fig_AmpliDelta}

	\includegraphics[width=0.40\textwidth]{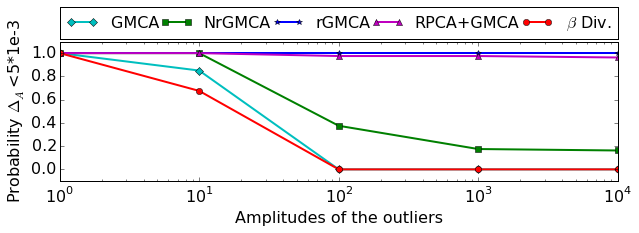}
	\caption{Probability of success versus the amplitude of the outliers.}
	\label{fig_probAmpli}
\end{figure}

We create $\nbreMesures=16$ measures of $\nbreSources=8$ sources with $\nbreSamples=1024$ samples to which a Gaussian noise is added. These sources are drawn from a Bernoulli-Gaussian law with parameter of activation $0.05$ and are scaled so that the largest entry is equal to 100. The entries of $\Amel$ are drawn from a centered normal law, and each column of $\Amel$ is then normalized. The positions of the outliers are such that: 160 are drawn uniformly at random and 10 columns are entirely corrupted. Their amplitudes are Gaussian with standard deviation chosen according to the axis of fig.\ref{fig_AmpliDelta}.

One can notice than the weighting procedure clearly enhances the estimation of $\Amel$: rGMCA outperforms NrGMCA, especially when the outliers are large fig.\ref{fig_AmpliDelta}. Contrary to the others methods, the combination PCP+GMCA and rGMCA are almost not influenced by the amplitudes of the outliers and are very likely to return good estimates of $\Amel$ fig.\ref{fig_probAmpli}.

\subsection*{Influence of the number of outliers}
\begin{figure}[h]
	\centering
	\includegraphics[width=0.40\textwidth]{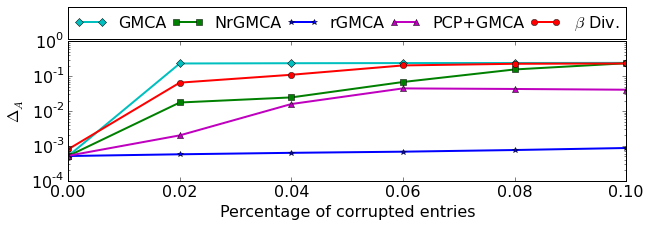}
	\caption{$\Delta_A$ versus the percentage of data corrupted by outliers.}
	\label{fig_NbreOutliers}

	\includegraphics[width=0.40\textwidth]{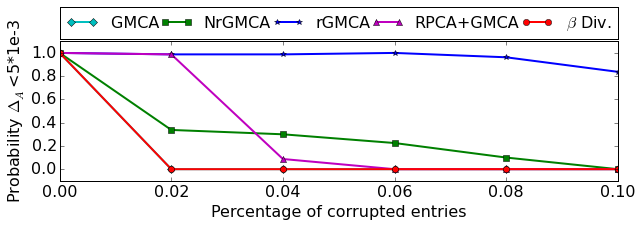}
	\caption{Probability of success versus the percentage of data corrupted by outliers.}
	\label{fig_probOutl}
\end{figure}

The components $\Amel$, $\Sources$ and $\Gaussien$ are similar to the ones presented in the previous subsection.
The percentage of grossly corrupted entries is fixed according to the axis of fig.\ref{fig_NbreOutliers}. Half of the outliers are drawn uniformly at random and the others correspond to the entirely corrupted columns.
Their amplitudes are Gaussian with a standard deviation of $100$.

All the methods are equivalent for $\Delta_A$ without outliers.  
The estimations of $\Amel$ obtained with PCP+GMCA, GMCA, NrGMCA and the $\beta$-divergence are hampered quite quickly by the percentage of corruption of the data fig.\ref{fig_NbreOutliers}. On the other side, rGMCA is robust while the percentage of corruption is moderated fig.\ref{fig_probOutl} .

\subsection*{NMR spectrum identification and influence of the number of observations}
\begin{figure}[h]
	\centering
	\includegraphics[width=0.40\textwidth]{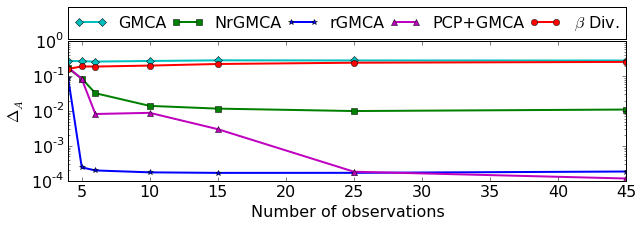}
	\caption{$\Delta_A$ versus the number of observations.}
	\label{fig_nbreM}

\end{figure}

In this subsection, we evaluate the algorithms in the field of the biomedical engineering with simulated data. We propose to separate the different Nuclear Magnetic Resonance spectra of a simulated mixture which can represent the data provided in NMR spectroscopy. By performing BSS on the mixture of spectra, we should be able to identify the different molecules of the mixture \cite{rapin2014nmf}. \\
The estimated NMR spectrum of the menthone, the folic acid, the ascorbic acid and the myo-inositol from SDBS \footnote{http://sdbs.db.aist.go.jp} are convolved with a Laplacian kernel of 2-samples width at half maximum (implementation from pyGMCA \footnote{http://www.cosmostat.org/software/gmcalab/}). The number of observations is set according to fig.\ref{fig_nbreM} . The Gaussian noise $\Gaussien$ is drawn from a Gaussian law with a standard deviation of $0.1$. The outliers are drawn from a Gaussian law with standard deviation $10^{3}$, so that $20$ columns and 1\% of the entries, broadly distributed, are corrupted. \\

\begin{figure}[h]
	\centering
	\includegraphics[width=0.24\textwidth]{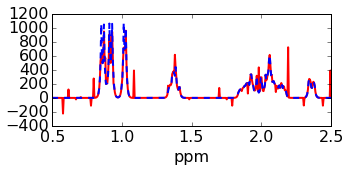}
	\includegraphics[width=0.24\textwidth]{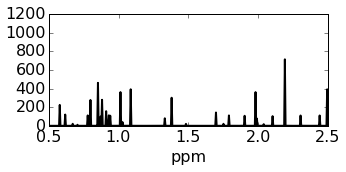}
	\includegraphics[width=0.24\textwidth]{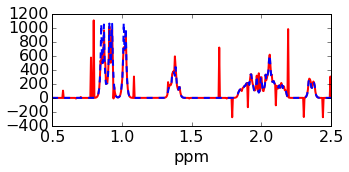}
	\includegraphics[width=0.24\textwidth]{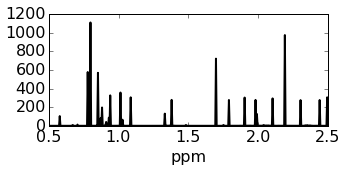}	
	
	\caption{Estimates of the menthone's NMR spectrum with rGMCA (top of the images) and PCP+GMCA (bottom of the images), with $m=6$. On the left: estimated spectrum in red and reference in blue-dashed lines - zoomed in on the support of the reference. On the right: magnitude of the difference between the estimate and the reference.}
	\label{fig_menthone}
\end{figure}

Despite the effectiveness of rGMCA and PCP+GMCA if $\nbreMesures\gg\nbreSources$ fig.\ref{fig_nbreM}, none of these algorithms are able to handle the outliers if $\nbreMesures=\nbreSources$.
The algorithm rGMCA is the only one that provides a correct estimate of $\Amel$ for $m\geq 5$ by means of the weighting scheme. However, rGMCA cannot clearly separate the original outliers from the estimated sources, fig.\ref{fig_menthone}.

\section{Conclusion}
We introduce a novel method to separate sparse sources in the presence of outliers. The proposed method relies on the joint sparsity-based separation of the outliers and the sources. This strategy allows us to implement a weighting scheme that penalizes corrupted data samples, which is shown to highly limit the impact of the outliers on the estimated sources. Numerical experiments demonstrate the good and consistent performances of our algorithm to robustly estimate the mixing matrix. Future work will focus on generalizing the proposed approach to enforce sparsity in a transformed domain.

\ifCLASSOPTIONcaptionsoff
  \newpage
\fi

\bibliographystyle{IEEEtran}

\bibliography{IEEEabrv,bib22}

\end{document}